\documentclass[aps,prb,reprint,superscriptaddress]{revtex4-2}

\usepackage{graphicx}
\usepackage{siunitx}
\usepackage{bbm}
\usepackage{amsmath,amsfonts,amssymb,braket,bbold,mathtools,array}
\usepackage[colorlinks,citecolor=blue,urlcolor=blue,bookmarks=false,hypertexnames=true]{hyperref} 

\newcommand{\eps}{\varepsilon}

\renewcommand{\vec}[1]{\boldsymbol{#1}}

\begin{document}

\title{Spin shuttling in a silicon double quantum dot}

\author{Florian Ginzel}
\affiliation{Department of Physics, University of Konstanz, D-78457 Konstanz, Germany}

\author{Adam R. Mills}
\author{Jason R. Petta}
\affiliation{Department of Physics, Princeton University, Princeton, New Jersey 08544, USA}

\author{Guido Burkard}
\affiliation{Department of Physics, University of Konstanz, D-78457 Konstanz, Germany}

\date{\today}

\begin{abstract}
The transport of quantum information between different nodes of a quantum device is among the challenging functionalities of a quantum processor.
In the context of spin qubits, this requirement can be met by coherent electron spin shuttling between semiconductor quantum dots.
Here we theoretically study a minimal version of spin shuttling between two quantum dots. To this end, we analyze the dynamics of an electron during a detuning sweep in a silicon double quantum dot (DQD) occupied by one electron. Possibilities and limitations of spin transport are investigated. %This research is motivated both by the demand for long and intermediate range interactions in quantum information devices and by recent experimental progress\cite{GaAsLine,Mills}. 
Spin-orbit interaction and the Zeeman effect in an inhomogeneous magnetic field play an important role for spin shuttling and are included in our model. Interactions that couple the position, spin and valley degrees of freedom open a number of avoided crossings in the spectrum allowing for diabatic transitions and interfering paths. The outcomes of single and repeated spin shuttling protocols are explored by means of numerical simulations and an approximate analytical model based on the solution of the Landau--Zener problem. We find that %high-fidelity spin-shuttling within a few nanoseconds 
a spin infidelity as low as $1-F_s\lesssim 0.002$ with a relatively fast level velocity of $\alpha = \SI{600}{\micro\electronvolt\per\nano\second}$ is feasible for optimal choices of parameters or by making use of constructive interference. 
\end{abstract}

\maketitle

\section{Introduction}

The spin of a single electron confined to a semiconductor quantum dot (QD) represents a highly coherent and controllable qubit realization for quantum information tasks
\cite{PhysRevA.57.120,RevModPhys.79.1217,doi:10.1146/annurev-conmatphys-030212-184248,Russ_2017}. A crucial ingredient for a quantum computer, however, is the interaction between arbitrary pairs of qubits within the device. Short-range interaction over distances on the order of $\SI{50}{\nano\meter}$ is mediated by the exchange interaction while long-range connectivity over $\si{\centi\meter}$ distances can be provided by spin-photon coupling \cite{PhysRevA.69.042302,PhysRevB.74.041307,Benito2017,Mi2018,Samkharadze1123,Landig2018,Benito2019}.

For the intermediate length-scale there are alternative approaches that do not require additional components such as a microwave cavity. Proposed solutions include information transfer between stationary qubits via a chain of exchange-coupled spins \cite{PhysRevLett.96.247206,PhysRevLett.98.230503,SWAP,Chan2020} or the transport of mobile qubits in a sliding potential well \cite{Shilton_1996,PhysRevB.88.075301,JadotSAW2020}. Adiabatic passage protocols \cite{PhysRevB.70.235317,PhysRevA.73.032319,CTAPexpPlatero,PhysRevLett.112.176803} are another approach that are currently of great interest \cite{Ban2019,Groenland2019,Stefanatos2019,GullansCTAP}.

A different flavor of mobile qubits are spins which are shuttled in a bucket-brigade manner between neighboring empty quantum dots \cite{Taylor2005}. This method to turn stationary into moving qubits has received much attention recently \cite{PhysRevA.96.012309,PhysRevB.97.245428,PhysRevB.99.155421,ZhaoHu2017,HuShuttling,ChargeNoise,cascade-vDiepen,Buonacorsi2020}. Coherent spin transfer has already been demonstrated in GaAs devices \cite{GaAsCircle,GaAsLine}, while charge shuttling down an array of 9 series-coupled QDs has been demonstrated in silicon \cite{Mills} and applications beyond transport are conceivable \cite{PhysRevB.86.245307,PhysRevB.88.245305}.

In bucket brigade shuttling, control of the QD gate voltages is used to drive the electron across a charge transition while avoiding hot spots where the spin relaxation rate is enhanced due to degeneracies between interacting spin and valley states \cite{PhysRevB.72.155410,PhysRevB.71.205324,PhysRevLett.110.196803}. A useful protocol must be robust against environmental effects \cite{PhaseRandomization,PhysRevLett.122.050501,ChargeNoise} and much faster than the relaxation and decoherence time of the spin, but at the same time slow enough to avoid errors due to non-adiabatic transitions between the (instantaneous) eigenstates \cite{PhysRevB.50.11902}. Realistically, the necessity of a trade-off between the spin transfer time and the shuttling fidelity can be anticipated.

The transport between neighboring QDs is affected by the spin-orbit interaction (SOI) that couples the spin of the electron to its momentum \cite{PhysRevLett.110.196803,PhysRevB.85.125312,NatCommun2.556,doi:10.1146/annurev-conmatphys-030212-184248}. This mechanism opens avoided crossings between opposite spin states, leading to spin-flip tunneling between neighboring QDs. In silicon the SOI is comparably weak but still relevant for quantum information tasks \cite{PhysRevB.98.245424,SOIMeasurement1,SOIMeasurement2}.

Another peculiarity of silicon-based QDs is the valley degree of freedom \cite{PhysRevLett.88.027903,PhysRevB.81.115324,Hollmann2019} with a two-dimensional, spin-like Hilbert space. The origin of the valley is the six-fold degenerate conduction band minimum in silicon which is partially lifted in a two-dimensional electron system
\cite{RevModPhys.54.437,PhysRevB.20.734,doi:10.1063/1.1637718,PhysRevB.80.081305,PhysRevB.84.155320}. 
%As a result, the six-fold degenerate minimum is split into a four-fold degenerate excited state and a two-fold degenerate ground state whose degeneracy is generally lifted in a perpendicular potential \cite{PhysRevB.20.734,doi:10.1063/1.1637718,PhysRevB.80.081305,PhysRevB.84.155320}. 
The valley splitting between the two lowest valley states, typically in the range of some $10-\SI{100}{\micro\electronvolt}$ \cite{PhysRevApplied.11.044063,Hollmann2019}, depends on the microscopic environment \cite{PhysRevB.75.115318,doi:10.1063/1.2387975,PhysRevB.19.3089,Yang2013ncomms,Hollmann2019}.

The theoretical framework to describe a driven two-level system with only one avoided crossing is the famous Landau--Zener (LZ) model \cite{AnnPhys210.16,PhysRep492.1,PhysRevA.53.4288,PhysRevLett.97.150502}. 
%which predicts diabatic transitions between the eigenstates of the system.
Extensions to the LZ model for multiple avoided crossings exist \cite{PhysRevB.56.13360,Demkov_1995,PhysRevA.75.013417,PhysRevA.56.232,PhysRevA.58.4293}, but it remains intrinsically challenging to characterize the error mechanisms limiting electron spin shuttling in a realistic solid state environment.
Further extensions to the LZ model known in the literature include different types of noise \cite{PhysRevA.91.052103,PhysRevB.67.144303,PhysRevB.76.024310,PhysRevB.87.165425,PhysRevB.87.224301}. Of particular interest for spin shuttling is $1/f$-noise. This so called charge noise stemming from electrical fluctuations \cite{doi:10.1063/1.4954700,PhysRevB.91.235411,PhysRevB.94.165411,Russ_2017,Hooge1997,Taylor2006} is the leading source of decoherence in a nuclear spin-free host material such as isotopically purified $^{28}$Si \cite{EnrichedSilicon}. It has been shown that charge noise can be a limiting factor for the shuttling fidelity \cite{ChargeNoise}.
Here, we model the coherent single-electron spin transfer in a tunnel-coupled silicon double quantum dot to understand the most elementary unit of any spin shuttling protocol and the underlying multilevel LZ physics. Our results show that even without environmental noise the shuttling fidelity can be severely limited by non-ideal system parameters. 

The remainder of this article is organized as follows. 
In Sec.~\ref{sec_Model}, a model for the spin and charge degrees of freedom of a single electron in a DQD in an inhomogeneous magnetic field and in the presence of SOI is derived.
Results for spin shuttling without (with) regard of the valley are presented in Sec.~\ref{sec_spin_shuttling} (Sec.~\ref{sec_Valley}). In particular, we discuss single shuttles in Sec.~\ref{sec_SingleShuttles} and repeated shuttling due to periodic driving with and without decoherence in Sec.~\ref{sec_SequentialShuttling}.
%The Hamiltonian describing spin shuttling in the presence of a (lifted) valley degeneracy is presented in Sec.~\ref{sec_Hvalley}, before the results will be discussed in Sec.~\ref{sec_ValleyChargeError}.
%In Sec.~\ref{sec_AnalyticalModel}, an analytical model based on the LZ formula %by obtaining an effective two-level Hamiltonian for each avoided crossing 
%is derived and evaluated.
Finally, our results are summarized in Sec.~\ref{sec_conclusion}.

\section{Model for charge and spin shuttling\label{sec_Model}}

\begin{figure}
\includegraphics[width=0.5\textwidth]{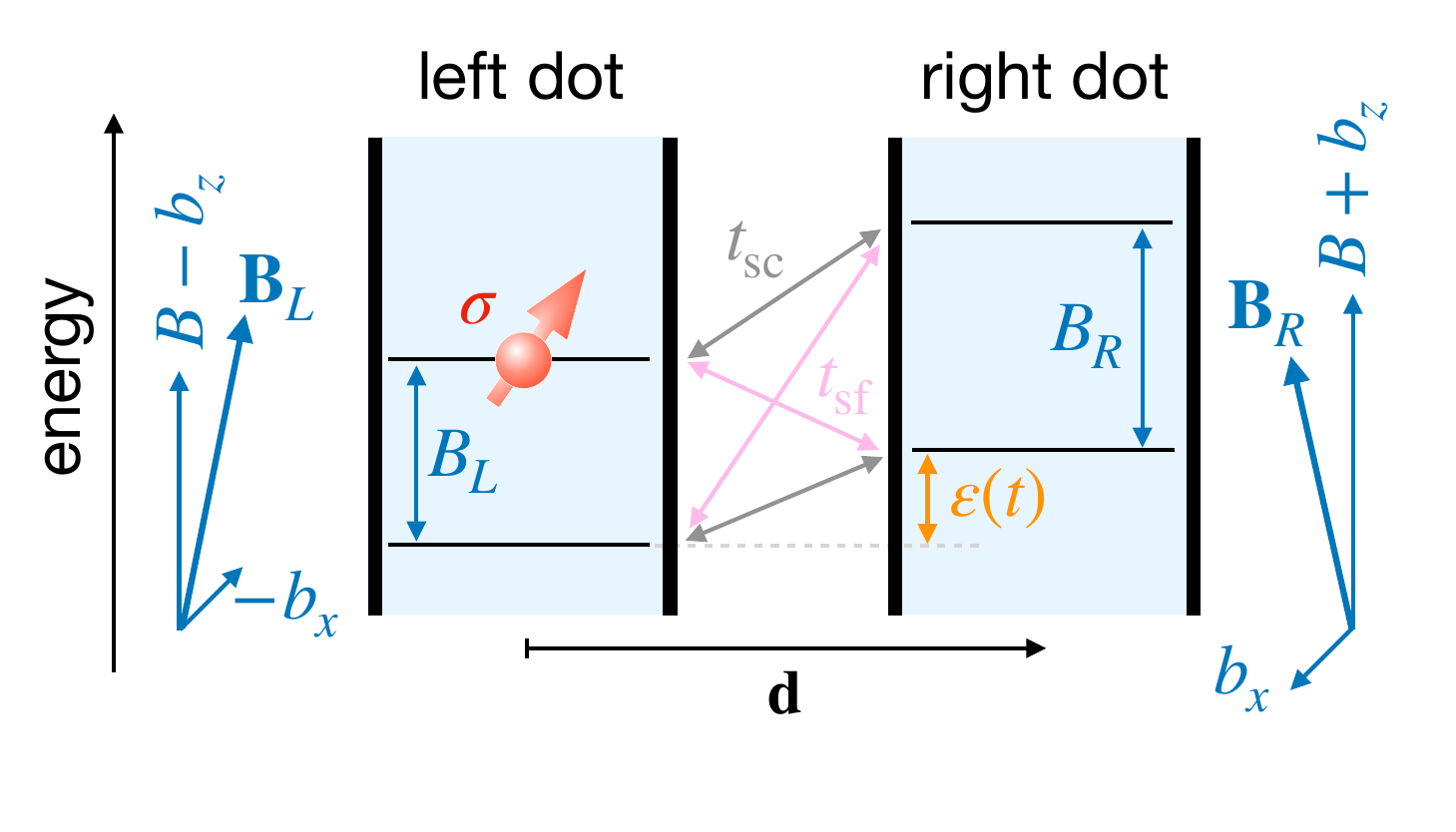}
\caption{Energy level diagram of the system under consideration: two quantum dots (QDs) filled with one electron with spin $\sigma$. The vector $\vec{d}$ points from the center of the left dot to the center of the right dot and thus determines the orientation of the DQD in the crystal. An electron is shuttled from the left QD to the empty right QD in the presence of a global magnetic field $\vec{B}$. A micromagnet can cause a magnetic field difference $2\vec{b}=2(b_x,b_y,b_z)$ between the left and right QD resulting in different total field 
$\vec{B}_{L,R}=\vec{B}\pm \vec{b}$ and Zeeman splittings $B_{L(R)}$. The time dependence of the level detuning $\eps(t)$ is chosen such that it conveys the electron from left to right. In addition to spin-conserving hopping $t_\mathrm{sc}$ a spin-flip tunneling term $t_\mathrm{sf}$ occurs due to the SOI and non-collinear magnetic fields in the two dots.\label{fig_system}}
\end{figure}

The minimal model for electron shuttling considered here is a double quantum dot (DQD) with Zeeman-split spin-levels in each dot, as depicted in the energy level diagram Fig.~\ref{fig_system}. Denoting the spin with the Pauli operators $\sigma_i$ and the position in the left-right basis with the Pauli operators $\tau_i$, where $i\in\{x,y,z\}$, the energy levels and the spin-conserving hopping between the dots are described by
\begin{equation}
H_0 = \frac{\varepsilon}{2} \tau_z \otimes \mathbbm{1} - t_c \tau_x \otimes \mathbbm{1} + \frac{B}{2} \mathbbm{1} \otimes \sigma_z.\label{eq_H0}
\end{equation}
Here, $\varepsilon=E_L-E_R$ denotes the energy detuning between the left and right dot states, which can be controlled by gate voltages \cite{PhysRevB.76.075306}. In the following sections, the shuttling protocol will consist of a detuning sweep $\varepsilon(t)$ (or repeated detuning sweeps) across the interdot charge transition \cite{Mills}. The inter-dot tunnel coupling is given by $t_c$. Additionally, a homogeneous magnetic field $\vec{B}=B\vec{\hat{z}}$ defining the $z$-axis is included, where the Zeeman splitting $B$ is given in energy units. For minimal disturbance an in-plane magnetic field parallel to the DQD-axis is favorable \cite{RevModPhys.79.1217}.% The Zeeman splitting $B$ contains the global magnetic field and the homogeneous part of a static hyperfine interaction in energy units.

To include magnetic field gradients, i.e., local differences of the Zeeman splitting and the inhomogeneous effects of a static hyperfine interaction, a term
\begin{equation}
H_\mathrm{grad} = \frac{1}{2} \tau_z \otimes  \vec{b}\cdot\vec{\sigma}\label{eq_Hgrad}
\end{equation}
is added. The difference in the Zeeman field between the two dots is given by the vector $2\vec{b} = 2(b_x,b_y,b_z)$, and $\vec{\sigma} = (\sigma_x,\sigma_y,\sigma_z)$ denotes the vector of the spin Pauli matrices. A detailed discussion of the different contributions to the magnetic field is given in Appendix~\ref{app_MagField}.

%Spin-flip tunnelling due to the spin-orbit interaction (SOI) is introduced with a term linear in momentum $\vec{p}$ \cite{PhysRevLett.110.196803,PhysRevB.85.125312,NatCommun2.556,doi:10.1146/annurev-conmatphys-030212-184248}. We use an orthonormalized basis for the left and right charge states constructed from the lowest Fock-Darwin state in each dot to calculate the matrix elements of $\vec{p}$ and find that the SOI-Hamiltonian can be expressed as
%\begin{equation}
%H_\mathrm{SOI} = \vec{p} A\vec{\sigma} = |p| \tau_y \otimes \vec{e}_d A\vec{\sigma},
%\end{equation}
%with a matrix $A$ which depends on the orientation of the quantum well in the crystal. The unit vector $\vec{e}_d$ determines the orientation of the DQD in the quantum well and thus the direction of motion of the electron during the shuttling protocol. In general, the matrix $A=A_R+A_D$ contains Rashba terms $\alpha_R$ and Dresselhaus-like terms $\beta_D$ due to interface inversion asymmetry \cite{RevModPhys.79.1217,PhysRevB.77.155328}.
Spin-flip tunneling due to the spin-orbit interaction (SOI) is introduced with a contribution that contains a Rashba term $\alpha_R (p_{x'} \sigma_{y'}-p_{y'}\sigma_{x'})$ \cite{RevModPhys.79.1217,doi:10.1146/annurev-conmatphys-030212-184248} and a Dresselhaus-like term $\beta_D (p_{x'}\sigma_{x'}-p_{y'}\sigma_{y'})$ due to interface inversion asymmetry \cite{RevModPhys.79.1217,PhysRevB.77.155328}. The confinement is chosen along $z'||[001]$. Here, $x',y',z'$ denote the crystallographic axes. Both SOI terms can be combined into the Hamiltonian $H_\mathrm{SOI} = \vec{p} A R\vec{\sigma}$ with a matrix $A=A_R+A_D$ that contains $\alpha_R$, $\beta_D$ and with a rotation $R$ with $(\sigma_{x'},\sigma_{y'},\sigma_{z'})=R(\sigma_x,\sigma_y,\sigma_z)$. We use an orthonormalized basis for the left and right charge states constructed from the lowest Fock-Darwin state in each dot to calculate the matrix elements of $\vec{p}$ and find that the SOI-Hamiltonian can be expressed as
\begin{equation}
H_\mathrm{SOI} = |p| \tau_y \otimes \vec{\hat{d}} A R\vec{\sigma},
\end{equation}
with the unit vector $\vec{\hat{d}}=\vec{d}/|d|$ where $\vec{d}$ is the vector connecting the centers of the two QDs (Fig.~\ref{fig_system}). Intra-dot effects of the SOI such as corrections to the $g$-factor \cite{RevModPhys.79.1217,PhysRevB.98.195314,NatCommun9.1768} can be incorporated into $H_\mathrm{grad}$ and $H_0$.

With the vector $\vec{\Omega}= |p| \vec{\hat{d}} A R = (\mathrm{Im}(a),\mathrm{Re}(a),s)$ the SOI-Hamiltonian finally reads \cite{PhysRevB.88.245305,PhysRevLett.110.196803,PhysRevLett.117.206802}
\begin{equation}
H_\mathrm{SOI} = \tau_y \otimes \vec{\Omega}\cdot\vec{\sigma}.\label{eq_HSOI}
\end{equation}
The form and strength of the spin-orbit-interaction strongly depend on the geometry of the DQD relative to the crystal lattice \cite{doi:10.1146/annurev-conmatphys-030212-184248,PhysRevLett.119.176807}: The parameters $a$ and $s$ depend on the angles between the axes $x$, $z$ and the axes $x'$, $z'$ via $R$ and on the orientation of the DQD-axis in the $x'$-$y'$-plane through $\vec{\hat{d}}$. For the chosen confinement direction $\alpha_R$ and $\beta_D$ enter via the SOI-matrix
\begin{equation}
A=\left(\begin{array}{ccc}
\beta_D & \alpha_R & 0 \\ -\alpha_R & -\beta_D & 0\\ 0 & 0 & 0
\end{array}\right).
\end{equation}
The shape of the QDs and the inter-dot distance determine the left and right electron wavefunctions and their overlap and thus impact the matrix element $|p|$. Note that the intra-dot terms depend on $\vec{B}$, the shape of the QDs, as well as the width of the quantum well \cite{PhysRevB.98.195314}. 
%Local and global corrections to the $g$-factor due to the SOI \cite{PhysRevB.98.195314,NatCommun9.1768} can be incorporated in the respective terms in $H_\mathrm{grad}$ and $H_0$.% if the $g$-factors are isotropic.

Theory and experimental results obtained from SiMOS platforms in Refs.~\cite{doi:10.1146/annurev-conmatphys-030212-184248,PhysRevB.98.245424,SOIMeasurement1,SOIMeasurement2} for a DQD along the $[110]$ crystal axis suggest an estimated range for the SOI-parameter $a\approx\sqrt{2}(i-1)(1\pm 0.4)\,\si{\micro\electronvolt}$ and $s=0$ for a typical interdot separation of $\SI{50}{\nano\meter}$. The SOI is kept constant at $a=\sqrt{2}(i-1)\,\si{\micro\electronvolt}$ throughout our analysis.

The total Hamiltonian is then the sum of all contributions,
\begin{equation}
H' = H_0(\eps) + H_\mathrm{grad} + H_\mathrm{SOI}.\label{eq_H-Original}
\end{equation}
%Diagonalizing $H'$ results in the instantaneous eigenstates $E_1(\eps)\leq E_2(\eps) \leq E_3(\eps) \leq E_4(\eps)$ depicted in Fig.~\ref{fig_spectrum}. The diabatic energy levels with the charge and spin configuration denoted by $(\sigma,0)$ [$(0,\sigma)$] for an electron with spin $\sigma$ in the left [right] and the instantaneous eigenstates are indicated in Fig.~\ref{fig_spectrum}. Avoided crossings between states $E_2$ and $E_3$ with opposite spin at $\varepsilon \approx \pm B$ are opened by the spin-flipping term $a$ from the SOI or the transverse magnetic gradient $b_{x(y)}$. The longitudinal magnetic gradient determines the position of the avoided crossings between states with opposite charge configuration (between $E_1$ and $E_2$ and between $E_3$ and $E_4$ in Fig.~\ref{fig_spectrum}) opened by $t_c$ at $\varepsilon \approx \pm b_z$.
The Hamiltonian $H'$ is defined with respect to a global basis where the spin is projected onto the same quantization axis determined by $\vec{B}$ in both QDs. To describe a spin shuttling experiment we assume that a basis of dot-localized eigenstates is used to prepare and measure the spin. We refer to this basis as local spin basis and derive it from the limit of isolated dots, $t_c=a=0$. In this limit the $2\times 2$ Hamiltonian of each dot can be diagonalized individually,
\begin{equation}
U_{L(R)} \left(\frac{B\mp b_z}{2}\,\sigma_z \mp\frac{b_x}{2}\, \sigma_x \pm\frac{\eps}{2}\right)
U_{L(R)}^\dagger
=
\frac{B_{L(R)}}{2}\, \sigma_z \pm\frac{\eps}{2}.\label{eq_lmbDef}
\end{equation}
For simplicity, the axes of the basis of $H'$ have been chosen such that $b_y=0$. The total magnetic field in left (right) dot is
\begin{equation}
B_{L(R)} =\sqrt{(B\mp b_z)^2+|b_x|^2}.
\end{equation}

\begin{figure}
\includegraphics{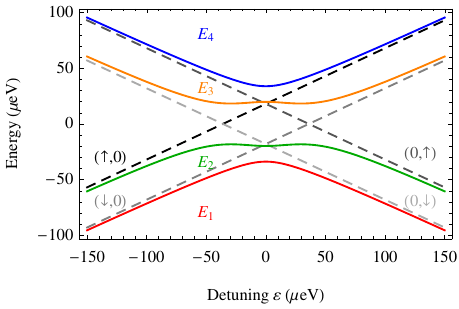}%
\caption{Spectrum of the Hamiltonian $H'$, Eq.~(\ref{eq_H-Original}), as a function of the detuning $\eps$, with diabatic (dashed) and adiabatic (solid) states. Spin-conserving tunneling $t_c$ opens avoided crossings at $\varepsilon \approx \pm b_z$ and the spin-flipping interactions $a$ and $\vec{b}$ open avoided crossings at $\varepsilon \approx \pm B$. Plot parameters are $t_c=\SI{21}{\micro\electronvolt}$, $B=\SI{30}{\micro\electronvolt}$, $\vec{b}=(\SI{30}{\micro\electronvolt},0,0)$ and $a=\sqrt{2}(i-1)\,\si{\micro\electronvolt}$. The spin ground state is labeled with $\downarrow$ and the excited spin state with $\uparrow$, respectively; by $(\sigma,0)$ [$(0,\sigma)$] we denote that an electron with spin $\sigma$ is localized in the left [right] dot.\label{fig_spectrum}}
\end{figure}

We define the Hamiltonian $H$ by the transformation $U = U_L \oplus U_R$,
\begin{equation}
H = UH'U^\dagger .\label{eq_H-lmb}\\
\end{equation}
In the limit $t_c=a=0$, $H$ is diagonal. In the following, we refer to the basis states of the frame defined by $U$ as diabatic states. The corresponding energy levels are plotted in Fig.~\ref{fig_spectrum} as dashed lines with the charge and spin configuration denoted by $(\sigma,0)$ [$(0,\sigma)$] for an electron with spin $\sigma$ in the left [right] dot. The spin ground state is labeled with $\sigma=\downarrow$ and the excited spin state with $\sigma =\uparrow$, respectively. The energy levels of states with opposite spin and charge configuration cross at $\eps=\pm (B_R+B_L)/2\approx \pm B$, the energies of states with same spin but opposite charge configuration cross at $\eps=\pm (B_R-B_L)/2\approx \pm b_z$.

Explicitly, the unitary $U_i$, $i=L,R$ is given by the rotation
\begin{equation}
U_i = \left(
\begin{array}{cc}
\cos\left(\vartheta_i/2\right) & \sin\left(\vartheta_i/2\right)\\
-\sin\left(\vartheta_i/2\right) & \cos\left(\vartheta_i/2\right)
\end{array}
\right).
\end{equation}
The angles $\vartheta_{i}/2$, can be understood as the angles between the global basis of $H'$ and the local spin basis of $H$. They are defined as the polar angle of the vector
\begin{equation}
\vec{u}_{L(R)}=\left(B_{L(R)}-(B\mp b_z), \pm b_x\right).
%v_{L(R)}=\left(\frac{B_{L(R)}-(B\mp b_z)}{b_x}, \pm 1\right).
\end{equation}

With finite inter-dot couplings the crossings of the diabatic states are opened to avoided crossings. The spin-conserving tunneling matrix element
\begin{eqnarray}
t_\mathrm{sc} &=& \cos\frac{\vartheta_L}{2} \left( (t_c+is) \cos\frac{\vartheta_R}{2}  +a \sin\frac{\vartheta_R}{2} \right) \nonumber\\
&& - \sin\frac{\vartheta_L}{2} \left( a^* \cos\frac{\vartheta_R}{2} -(t_c -is)\sin\frac{\vartheta_R}{2} \right)\label{eq_c}
\end{eqnarray}
opens the crossings at $\eps \approx \pm b_z$. The crossings at $\eps\approx\pm B$ are opened by the spin-flip tunneling matrix element
\begin{eqnarray}
t_\mathrm{sf} &=& \cos\frac{\vartheta_L}{2} \left( a\cos\frac{\vartheta_R}{2}  - (t_c +is) \sin\frac{\vartheta_R}{2} \right) \nonumber\\
&& + \sin\frac{\vartheta_L}{2} \left( (t_c-is) \cos\frac{\vartheta_R}{2} +a^*\sin\frac{\vartheta_R}{2}  \right). \label{eq_f}
\end{eqnarray}
Although $b_x$ does not couple different dots it does lead to a dot-dependent tilting $\vartheta_{L(R)}$ of the spin orientation which in turn affects the tunneling matrix elements. In view of spin shuttling this is crucial since the probabilities for spin conserving and spin-flip charge transitions estimated from the LZ formula are exponentially sensitive on $|t_\mathrm{sc}|^2$, $|t_\mathrm{sf}|^2$ \cite{AnnPhys210.16,PhysRep492.1}. In the limit $b_x\to 0$ the local bases align, $\vartheta_L-\vartheta_R\to 0$, thus the spin-flip tunneling $t_\mathrm{sf}$ is only due to the SOI.
The energy levels of the instantaneous eigenstates $E_1(\eps)\leq E_2(\eps) \leq E_3(\eps) \leq E_4(\eps)$ of $H$ are depicted in Fig.~\ref{fig_spectrum} as solid lines.

The shuttling protocol is chosen to be a linear detuning ramp from $-\eps_0$ to $+\eps_0$ with level velocity $\alpha$ within a time interval $0\leq t \leq 2\eps_0/\alpha$,
\begin{equation}
\eps(t) = \alpha t - \eps_0.\label{eq_protocolstart}
\end{equation}
The tunnel coupling $t_c$ and all magnetic fields are kept constant during the protocol.

The choice of a linear detuning ramp is motivated by both recent experiments \cite{Mills,UNSW} and the fact that by this choice some analytic estimations can be obtained from the LZ model (Sec.~\ref{sec_AnalyticalModel}). Reducing the level velocity in the vicinity of the avoided crossings will reduce the probability of LZ transitions but also requires more time during which the spin state can suffer from noise. We expect that fidelity and protocol duration can be further improved by optimizing the protocol \cite{PhysRevLett.116.230503,PhysRevX.7.011021}. However, this is beyond the scope of this work. Another common protocol choice are constant-adiabaticity pulses for whose high fidelity shuttling has been proposed recently \cite{Buonacorsi2020}.

To evaluate the shuttling protocol it is assumed that initially a single electron is prepared in the left dot at the beginning of the ramp, $|\mathrm{in}\rangle = |\sigma,0\rangle$. At time $t_\mathrm{end}=2\eps_0/\alpha$ the state of the system has evolved to $|\mathrm{out}\rangle$. Since the aim of the protocol is an error-free spin transfer between the dots the fidelity \cite{NielsenChuang} $F_s=|\langle 0,\sigma|\mathrm{out}\rangle|^2$ is a measure for the success for the spin shuttling protocol. In general, $F_s$ depends on the spin $\sigma$ of the input state.% Note that the spin fidelity $F_s$ is bounded by the charge fidelity $F_c= \sum_{\sigma'} |\langle 0,\sigma'|\mathrm{out}\rangle |^2$ which measures the probability of faithful charge transport.

\section{Spin shuttling\label{sec_spin_shuttling}}

In this section we numerically solve the problem of spin shuttling. In Sec.~\ref{sec_SingleShuttles} only a single passage through the avoided crossing region is considered while in Sec.~\ref{sec_SequentialShuttling} a sequence of back and forth shuttling is analyzed. To determine a lower bound for the spin shuttling infidelity $1-F_s$ we numerically integrate the Schr\"odinger equation with degenerate spin levels during a finite-time detuning sweep and compute the charge infidelity $1-F_c= \sum_{\sigma'} |\langle 0,\sigma'|\mathrm{out}\rangle |^2$ where $F_c$ measures the probability of faithful charge transport.
Based on the findings for charge shuttling the tunneling is set to the fixed value $t_c=\SI{21}{\micro\electronvolt}$ and the level velocity is set to $\alpha = \SI{600}{\micro\electronvolt\per\nano\second}$ for the entire analysis. In the absence of spin and magnetic fields this choice allows a charge transport infidelity of $1-F_c\approx 10^{-5}$. Our choice of $t_c$ and $\alpha$ is based on recent experiments \cite{Mills}. Note that with this choice of the parameter $\alpha$, a shuttling protocol with $\eps_0$ in the order of $\si{\milli\electronvolt}$ can be completed multiple times within tens of nanoseconds, several orders of magnitude faster than the spin dephasing time of $T_2^*\gtrsim \SI{100}{\micro\second}$ observed in isotopically purified silicon \cite{Si28T2star}. 
In a nuclear spin-free host material such as isotopically purified $^{28}$Si the time scale of $T_2^*$ is set by charge noise %– electrical fluctuations with a $1/f$ power spectral density
\cite{doi:10.1063/1.4954700,PhysRevB.91.235411,PhysRevB.94.165411,Russ_2017,Hooge1997,Taylor2006}.% Charge noise is the leading source of decoherence in a nuclear spin-free host material such as isotopically purified $^{28}$Si \cite{EnrichedSilicon}.

\subsection{Single shuttles\label{sec_SingleShuttles}}

\begin{figure}[h!]
\includegraphics{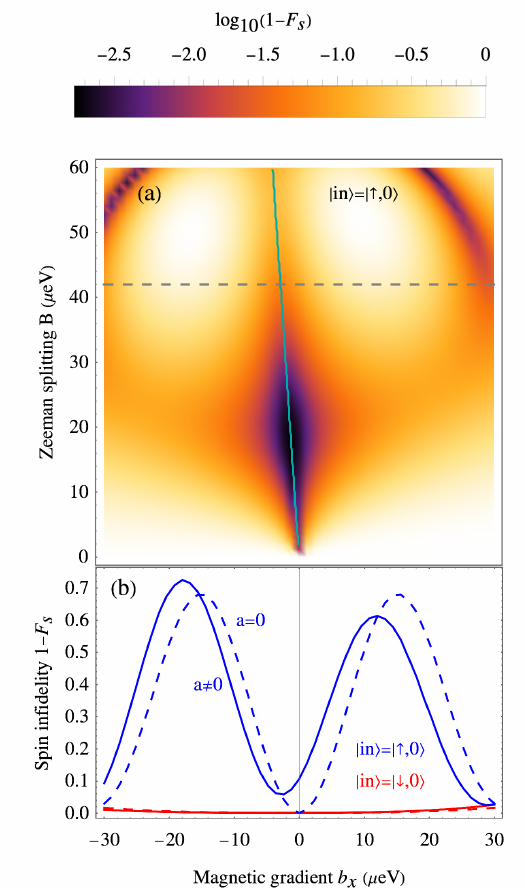}
\caption{Spin shuttling infidelity without valley degeneracy. (a) Logarithm of the spin shuttling infidelity $1-F_s$ for initialization in the excited spin state, $|\mathrm{in}\rangle = |\!\uparrow,0\rangle$, as a function of the transverse magnetic field difference $b_x$ and the Zeeman splitting $B$.
Free parameters are chosen as $t_c=\SI{21}{\micro\electronvolt}$ $\eps_0=\SI{8}{\milli\electronvolt}$ and $b_z=0$.
With increasing $b_x$ the spin-flip tunneling $t_\mathrm{sf}$ leads to diabatic transitions, and $1-F_s$ shows local maxima (minima) due to destructive (constructive) LZ interference. % between different paths.
The cyan line indicates the minimal $|t_\mathrm{sf}|$ for each $B$.
(b) Spin infidelity $1-F_s$ along a cut through panel (a) indicated by the dashed line ($B=2t_c=\SI{42}{\micro\electronvolt}$) for the excited spin state $|\mathrm{in}\rangle = |\!\uparrow,0\rangle$ (blue, solid) and the ground state, $|\mathrm{in}\rangle = |\!\downarrow,0\rangle$ (red, solid). The dashed lines correspond to the case without SOI, $a=0$. The ground state does not show interference in a single passage.\label{fig_SpinFids}}
\end{figure}

We numerically integrate the time-dependent Schr\"odinger equation
$i \hbar \partial_t |\psi(t)\rangle = H (t) |\psi(t)\rangle$
and plot the infidelity $1-F_s$ as a function of the magnetic field $B$ and gradient field $b_x$ in Fig.~\ref{fig_SpinFids}. The transverse magnetic field differences $b_x$ and $b_y$ have equivalent effects, thus, for simplicity $b_y = 0$ is chosen. In Fig.~\ref{fig_SpinFids}a, where the initial state is chosen to be the excited spin state $|\mathrm{in}\rangle = |\!\uparrow,0\rangle$, we observe two dominant features of the spin shuttling protocol: an increase of infidelity with increasing gradient $b_x$ and local extrema occurring for $B>2t_c$ due to interference between the probability to cross the charge transition either adiabatically in $E_2$ or involving diabatic transitions between $E_2$ and $E_3$. Figure \ref{fig_SpinFids}b shows a cut along the $b_x$-axis which highlights the effect of SOI and also shows the infidelity for the case of initialization in the ground state  $|\mathrm{in}\rangle = |\!\downarrow,0\rangle$.

The increase of infidelity due to increasing $b_x$ visible in Fig.~\ref{fig_SpinFids} for both input states can be explained by the fact that a non-vanishing transverse gradient $b_x \neq 0$ causes the local spin bases to be non-collinear. Consequently, a spin-flip tunneling term $t_\mathrm{sf}$, Eq.~(\ref{eq_f}), occurs even in the absence of SOI ($a=0$). The different spin projections in the left and right dot lead to an increase of spin infidelity due to diabatic transitions between states with opposite local spin eigenstates.

If $b_x\neq 0$ the SOI $a$ also contributes to the spin-conserving hopping term $t_\mathrm{sc}$, Eq.~(\ref{eq_c}), and can thus compensate the increase of infidelity to some extent. This is shown in Fig.~\ref{fig_SpinFids}b. The solid lines with SOI $a\neq 0$ are laterally shifted compared to the dashed curves with $a=0$, in particular, the minimum of $1-F_s$ coming from collinear quantization axes at $t_\mathrm{sf}=0$ occurs at finite $b_x$. As a result, on one flank of the dip the infidelity with $a\neq 0$ is smaller than with $a=0$ while on the opposite flank the infidelity is increased due to the combined effects of magnetic gradient and SOI.
The magnitude of the lateral shift in the $b_x$-$B$-plane is approximately proportional to $\mathrm{Re}(a)=|a|\cos(\mathrm{arg}\: a)$. Note that while in the cut in Fig.~\ref{fig_SpinFids}b the minimal $1-F_s$ for $|\mathrm{in}\rangle = |\!\uparrow,0\rangle$ is significantly increased by the presence of SOI ($a\neq 0$) the minimum of $1-F_s$ in the entire $b_x$-$B$-plane is reduced only by $\approx 0.1\,\%$ for our choice of $a$.

The local minima and maxima on both sides of the line with $t_\mathrm{sf}=0$ visible in $1-F_s$ with $|\mathrm{in}\rangle = |\!\uparrow,0\rangle$ are another effect of $t_\mathrm{sf}$. The spin-flip tunneling opens avoided crossings between the states with spin and charge configuration $(\uparrow,0)$ and $(0,\downarrow)$ as well as between $(\downarrow,0)$ and $(0,\uparrow)$. Thus, two paths can lead to faithful shuttling of the excited spin state, either adiabatically following $E_2$ or by a diabatic transition to $E_3$ followed by another diabatic transition back to $E_2$. The interference between the probability amplitudes of the two paths can lead to local maxima of $1-F_c$ and consequently $1-F_s$, which we call destructive interference, and local minima of $1-F_s$ which we deem constructive interference.
Interference extrema of first order are visible in Fig.~\ref{fig_SpinFids}a. In the vicinity of the maxima of $1-F_s$ one diabatic transition $E_2\to E_3$ is more likely than an adiabatic trajectory along $E_2$ or the successive transitions $E_2\to E_3 \to E_2$. This corresponds to a transition $(\uparrow,0)\to (\downarrow,0)$ into the excited charge state with a spin flip rather than a spin-conserving charge transition. Transfer of the spin ground state is not affected by the additional avoided crossings, as Fig.~\ref{fig_SpinFids}b emphasizes.

The longitudinal magnetic field difference $b_z$ has an effect only in a protocol with short ramp, i.e. small $\eps_0$. Since the avoided crossings opened by $t_c$ appear near $\eps \approx\pm b_z$ the length of the second part of the ramp after the anticrossing is changed. Thus, the phase of finite-time LZ oscillations \cite{PhysRevA.53.4288,FiniteTimeLZExp} relative to the end of the protocol is shifted. Consequently, the infidelity at $\eps(t) = \eps_0$ shows oscillations as a function of $b_z$. During a sufficiently long ramp $\eps_0 \gg \mathrm{max}(B,t_\mathrm{sc},t_\mathrm{sf})$ finite-time oscillations decay and become irrelevant.

To optimize the shuttling results, $t_c$ and $b_x$ should be chosen in a way that minimizes $t_\mathrm{sf}$ and maximizes $t_\mathrm{sc}$ to increase the probability of adiabatic electron transport. The loss of fidelity due to LZ interference can be avoided by either using a sufficiently weak magnetic field $B<2t_c$ or by tuning $B$ to exploit constructive interference. Furthermore, a long ramp $\eps_0 \gg \mathrm{max}(B,t_\mathrm{sc},t_\mathrm{sf})$ helps to avoid timing-related effects.

%\begin{figure}
%\includegraphics{Fig8.pdf}
%\caption{Shuttling of superposition states with $|\mathrm{in}\rangle = |+,0\rangle = (|\!\downarrow,0\rangle +|\!\uparrow,0\rangle)/\sqrt{2}$ for $B=\SI{42}{\micro\electronvolt}$, $\eps_0=\SI{0.8}{\milli\electronvolt}$ and $b_z=0$ (cf. Fig.~\ref{fig_SpinFids}b). The plot shows the overlap of $|\mathrm{out}\rangle$ with $|0,\uparrow\rangle$ (red), $|0,\downarrow\rangle$ (blue) and $|0,+\rangle$ (yellow). The final state oscillates between the even and odd spin superposition and the spin populations interfere.\label{fig_superpositions}}
%\end{figure}

In general, we will be interested in transporting general quantum states of the spin, rather than local spin eigenstates.
Spin superposition states are non-stationary in the chosen basis. When residing in dot $j$ the initial state $|\psi(0)\rangle = c_1|\!\downarrow\rangle + c_2|\!\uparrow\rangle$ evolves to the state $|\psi(0)\rangle = c_1 e^{i B_j t/2}|\!\downarrow\rangle + c_2 e^{-iB_j t/2}|\!\uparrow\rangle$. This oscillatory behaviour leads to a relative phase of the final superposition state. The outcome obtained from shuttling the basis states $|\pm,0\rangle = \frac{1}{\sqrt{2}}\left(|\!\downarrow,0\rangle \pm|\!\uparrow,0\rangle\right)$ oscillates between $|\mathrm{out}\rangle = |0,+\rangle$ and $|\mathrm{out}\rangle = |0,-\rangle$ as a function of the duration of the protocol and the local magnetic fields $B_{L(R)}$. In the vicinity of the destructive interference described in Fig.~\ref{fig_SpinFids} the probability for faithful transport of a spin superposition drops since at this point the component with $\sigma= \uparrow$ is not shuttled at all with high probability.

Beyond the effects known from the shuttling of states with binary spin, $|\mathrm{in}\rangle = |\sigma,0\rangle$, $\sigma\in\{\uparrow,\downarrow\}$, the projection of the final superposition state on the local spin eigenstates, $|\langle 0,\uparrow\!(\downarrow)|\mathrm{out}\rangle|^2$, shows an oscillatory pattern. This can be explained by the fact that if $|\mathrm{in}\rangle$ is a superposition state the two lowest-lying eigenstates $E_1$ and $E_2$ both have a finite population. For $t_\mathrm{sf}\neq 0$ there is a probability for diabatic transitions between them in the avoided crossing at $\eps_c = (B_L-B_R)/2\approx - b_z$.

Assuming widely spaced anticrossings we can apply the LZ formula to approximate the population of the eigenstates $E_1(\eps)$ and $E_2(\eps)$ directly after the avoided crossing at $\eps_c$. We assume the state before the anticrossing is $c_1(\eps_c-)|\!\downarrow\rangle + c_2 (\eps_c-)|\!\uparrow\rangle$ with amplitudes $c_i(\eps_c-) =\lim_{\eps\to (\eps_c- )} c_i(\eps)$ and phases $\varphi_i = \frac{1}{\alpha}\int_{-\eps_0}
^{\eps_c-} \mathrm{d}\eps\, E_i$, where $c_i(\eps_c\pm) =\lim_{\eps\to (\eps_c\pm 0)} c_i(\eps)$ are the limits from above and below. Then the coefficients evolve to \cite{AnnPhys210.16,PhysRep492.1}
%\begin{footnotesize}
%\begin{equation}
\begin{eqnarray}
\left(\begin{array}{c}
c_1  (\eps_{c}+) \\ c_2 ( \eps_{c}+)
\end{array}\right)
&=&
\left(\begin{array}{cc}
\sqrt{1-P}e^{-i\varphi_s} & -\sqrt{P} \\ \sqrt{P}  & \sqrt{1-P}e^{i\varphi_s} 
\end{array}\right)\nonumber \\
& & \times \left(\begin{array}{c}
|c_1(\eps_{c}-) | e^{i \varphi_1} \\ |c_2( \eps_{c}-) | e^{i \varphi_2}
\end{array}\right).
\end{eqnarray}
%\end{equation}
%\end{footnotesize}
Here, %$c_i(\eps_c\pm) =\lim_{\eps\to (\eps_c\pm 0)} c_i(\eps)$ are the limits from above and below, 
$P$ is the probability for a diabatic transition calculated from the LZ formula and $\varphi_s$ is the Stokes phase associated with the avoided crossing. This leads to the emergence of an interference term $\propto\cos(\varphi_1+\varphi_2+\varphi_s)$ in $|\langle 0,\downarrow | \mathrm{out}\rangle|^2$.
Note that in this estimation $|c_{1(2)}(\eps_{c}-)|^2$ are not equal to the initial populations since the avoided crossing opened by $t_\mathrm{sf}$ at $\eps = -(B_L+B_R)/2\approx - B$ has to be taken into account.

In more complex systems the loss of fidelity due to destructive LZ interference can be reduced by device optimization. A minimal example is a cyclic round trip in a triple quantum dot \cite{doi:10.1063/1.3258663,PhysRevB.82.075304,GaAsCircle} in triangular arrangement. Applying our model of spin shuttling it can be shown that by manipulating the complex phases of the tunneling matrix elements it is possible to engineer the phase shift during the charge transition.

\subsection{Sequential shuttling\label{sec_SequentialShuttling}}

\begin{figure}
\includegraphics{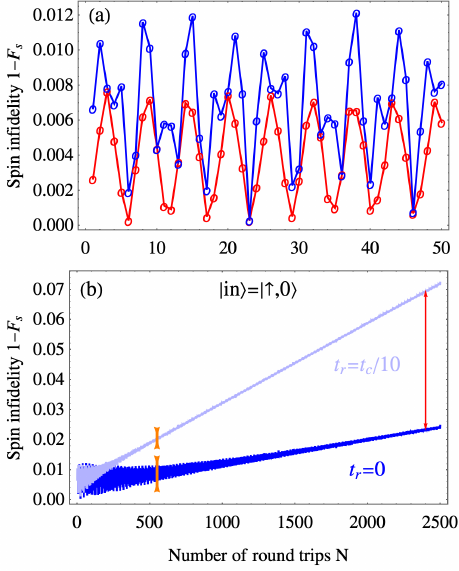}
\caption{(a) Spin infidelity $1-F_s$ for $|\mathrm{in}\rangle = |\!\uparrow,0\rangle$ (blue) and $|\mathrm{in}\rangle = |\!\downarrow,0\rangle$ (red) as a function of the number of round trips $N$ in the DQD, where $B = \SI{25}{\micro\electronvolt}$, $b_x = -\SI{3.4}{\micro\electronvolt}$, $b_z = \SI{1.56}{\micro\electronvolt}$, $t_c = \SI{21}{\micro\electronvolt}$ and $\eps_0=\SI{0.8}{\milli\electronvolt}$. The interference pattern in the excited state is more complex since the multiple anticrossings passed per transition give rise to several oscillating terms.
(b) Repeated shuttling of $|\mathrm{in}\rangle = |\!\uparrow,0\rangle$. The dark blue curve is the same as in (a), the light blue curve additionally takes into account the coupling to reservoirs described by Eq.~(\ref{eq_ME-res}) with $t_r=t_c/10$ and Coulomb repulsion between the neighboring dots $U_c=\SI{5}{\milli\electronvolt}$ and temperature $T = \SI{0.1}{\kelvin}$. The effects of spin-flip cotunneling (red) and decoherence (orange) are indicated with arrows. The $(1,0)\leftrightarrow(0,1)$ charge transition is crossed in the middle between the triple points involving the $(0,0)$ and $(1,1)$ regimes.\label{fig_sequence}}
\end{figure}

To access the infidelity more easily than in single shuttles, the electron can be shuttled back and forth between the dots $N$ times. At the end of the first ramp the reverse protocol is applied to complete the round trip. This cycle is repeated $N$ times. For $\eps_0=\SI{800}{\micro\electronvolt}$ and level velocity $\alpha=\SI{600}{\micro\electronvolt\per\nano\second}$ the time per round trip is $\SI{5.3}{\nano\second}$. The intrinsic spin relaxation with typical lifetimes $T_1$ in the order of $\si{\milli\second}$ to $\si{\second}$ \cite{PhysRevLett.104.096801,PhysRevLett.106.156804,YangLifetime2013,PhysRevLett.121.076801} can be neglected even for a long sequence with $\mathcal{O}\left(10^4\right)$ round trips with local eigenstates as initial states. The increase of the spin infidelity as a function of $N$, shown in Fig.~\ref{fig_sequence}, is thus predominantly due to the error mechanisms discussed in Sec.~\ref{sec_SingleShuttles}. In general, with a superposition state as initial state, the decoherence time $T_2$ has to be taken into account.
As shown in Fig.~\ref{fig_sequence}a, interference can also be observed with a spin initialized in the ground state $|\mathrm{in}\rangle=|\!\downarrow,0\rangle$ and then undergoing several shuttling round trips. This is a consequence of the system being swept though the same avoided crossing region multiple times, analogous to Landau--Zener-St\"uckelberg interferometry \cite{AnnPhys210.16,PhysRep492.1,HuShuttling}. For an electron in the excited spin state, interfering paths are available even for a single shuttling sweep, and thus the oscillations for $\sigma= \uparrow$ are the result of a superposition of multiple interference terms.

In a long sequence of shuttles the decay of fidelity is approximately modeled by the rate equation
\begin{equation}
\frac{\mathrm{d}}{\mathrm{d}N}\,\vec{n}(N) = \left(
\begin{array}{cccc}
 -c_1 & c_2 & 0 & 0 \\
 c_1 & -c_2-c_3 & c_4 & 0 \\
 0 & c_3 & -c_4-c_5 & c_6 \\
 0 & 0 & c_5 & -c_6 \\
\end{array}
\right) \vec{n}(N)\label{eq_RateEqSeq}
\end{equation}
with $\vec{n}(N)=\left[n_1(N),n_2(N),n_3(N),n_4(N)\right]$ the vector of populations of the four states. The rate equation describes four coupled levels with population $n_i$ where transitions can occur between level $i$ and the levels $i\pm 1$ adjacent in energy during each round trip. The asymptotic limit for any input state is $F_s=1/4$ with the population equally distributed between all four basis states. 

The interference due to sequential passage through the same avoided crossing described above can be suppressed due to charge decoherence associated with loss of the electron due to coupling to the source/drain reservoirs of the DQD (e.g. inelastic tunneling to the (0,0) or (1,1) charge state). To examine the effects of decoherence the interaction of each of the two QDs with one fermionic reservoir constituted of a two-dimensional electron gas (2DEG) is included. For example, the DQD's source/drain contacts can form such reservoirs. The reservoirs are coupled to the DQD by incoherent tunneling which does not conserve the DQD charge, introducing the charge states $(0,0)$ and $(1,1)$. Charge states with a doubly occupied quantum dot are neglected by assuming a large on-site Coulomb repulsion. The time evolution is then described by a Lindblad-form master equation (ME) which can be brought into the  form \cite{DensMatTheoApp}
\begin{equation}
\frac{\mathrm{d}}{\mathrm{d} t}\,\rho_{nm} = \frac{1}{i}[H,\rho]_{nm}+\delta_{nm}\sum_l w_{nl}\rho_{ll}-\gamma_{nm}\rho_{nm}.\label{eq_ME-res}
\end{equation}
The transitions rates are derived from Fermi's golden rule,
\begin{equation}
w_{mn}= 2\pi |t_r|^2 D n_F
\end{equation}
for an electron tunneling from one of the reservoirs to one of the dots and
\begin{equation}
w_{mn}= 2\pi |t_r|^2 D (1-n_F)
\end{equation}
for an electron tunneling to the reservoirs with the tunneling matrix element $t_r$ between a QD and the attached reservoir. The density of states of the 2DEG near the Fermi energy is given by $D$ and $n_F$ is the Fermi-Dirac distribution function evaluated at the energy of the added or removed electron. To doubly occupy the DQD the Coulomb energy $U_c$ between the QDs must be overcome. The definitions of the decoherence rates $\gamma_{nm}$ are given in Appendix~\ref{app_ME}.

The interaction with the reservoirs is negligible for a small number of shuttles, however, it can significantly impact the result of a long sequence in two ways, as Fig.~\ref{fig_sequence}b shows. A spin-flip cotunneling process between the dots and the reservoir which randomizes the spin in the DQD raises the infidelity. Additionally, due to decoherence, the oscillations caused by interference are damped as the incoherent tunneling is introduced.

\section{Spin and valley\label{sec_Valley}}

The Hamiltonian $H$ from Eq.~(\ref{eq_H-lmb}) does not take into account the valley degree of freedom \cite{PhysRevLett.88.027903,PhysRevB.81.115324,Hollmann2019}. Thus, the previous analysis applies to the limit where the valley degree of freedom does not affect the system dynamics, e.g. because the valley splitting exceeds all relevant energy scales appearing in the shuttling process, and to the case of systems without valley, e.g. quantum dots in GaAs or InAs. However, the valley in silicon cannot be neglected when the valley and Zeeman splittings are comparable. To analyze the effects of valley transitions in addition to the spin and orbital degrees of freedom, we extend our  model to a Hamiltonian $H_v$ acting on the product Hilbert space of charge, spin and local valley degrees of freedom.

\subsection{Valley Hamiltonian\label{sec_Hvalley}}

The general valley Hamiltonian for QD $i\in\{ L,R \}$ is given by \cite{PhysRevB.94.195305}
\begin{equation}
H_{\mathrm{valley},i} = \vec{v}_i\cdot\vec{\nu},\label{eq_HValleyGen}
\end{equation}
where $\vec{\nu}$ is the vector of Pauli operators for the valley degree of freedom and $\vec{v}_i$ is a vector that determines orientation and modulus of the valley splitting in dot $i$ with respect to a global valley basis. We then introduce the DQD Hamiltonian $H'_v$ for spin, position and valley in its global basis. Using the Hamiltonian $H$ from Eq.~(\ref{eq_H-lmb}) we define
\begin{equation}
H'_v = H  + \sum_{i\in\{L,R\}} H_{\mathrm{valley},i}.
\end{equation}

In analogy to the local spin eigenbasis, Eq.~(\ref{eq_lmbDef}), a unitary transformation $H_v= U_v H'_v U_v^\dagger$ is applied to diagonalize the valley Hamiltonian in each dot individually. The transformed Hamiltonian in the local valley eigenbasis has the form
\begin{eqnarray}
H_v =&& \sum_{i\in\{L,R\}} \left[ H_{ii} + \left(\begin{array}{cc}
E_{v,i} & 0 \\
0 & 0
\end{array}\right)\right] \nonumber\\
&& + \left[
H_{LR} \left(\begin{array}{cc}
\cos \vartheta & -\sin \vartheta \\
\sin \vartheta &\cos \vartheta
\end{array}\right) +\mathrm{h.c.} \right]
\end{eqnarray}
where $E_{v,i}$ denotes the valley splitting in QD $i$ and $\mathrm{h.c.}$ denotes the Hermitian conjugate. The Hamiltonian $H$ for spin and charge was divided into a tunneling contribution $H_{LR}+H_{RL}=\sum_{\sigma,\sigma'} \langle 0,\sigma' |H|\sigma,0\rangle |0,\sigma' \rangle \langle \sigma,0| +\mathrm{h.c.}$ and an intra-dot contribution $H_{LL}+H_{RR}=\sum_\sigma \langle \sigma,0|H|\sigma,0\rangle |\sigma,0\rangle\langle \sigma,0| + (0\leftrightarrow \sigma)$.
The angle $\vartheta = (\vartheta_L^v - \vartheta_R^v)/2$ with $0\leq\vartheta \leq \pi$ can be understood as the angle between the valley pseudospins in the QDs while $\vartheta_i^v$ is the angle between the local valley eigenbasis of dot $i$ and the global valley basis. Note that only the tunneling terms that couple the two dots depend on $\vartheta$.
\begin{figure}
\includegraphics{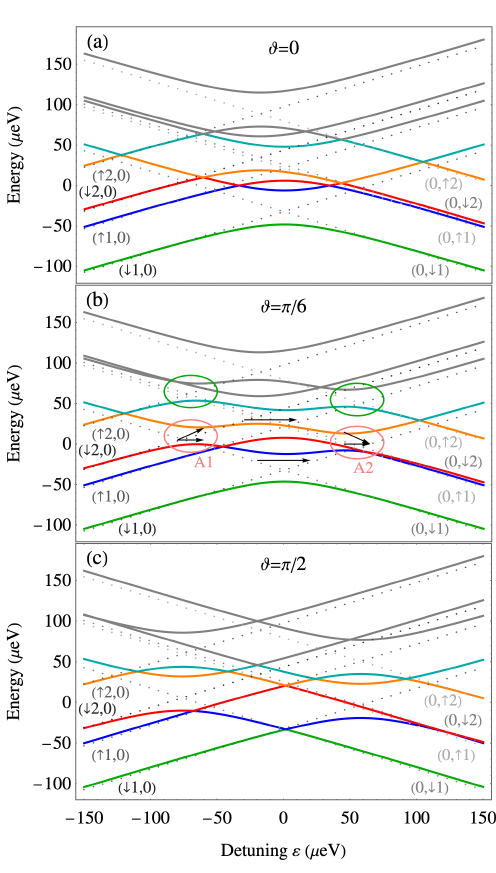}%
\caption{Spectrum of $H_v$ as a function of the detuning $\eps$ (solid) and diabatic basis states (dashed) for $B=\SI{54}{\micro\electronvolt}$, $t_c=\SI{21}{\micro\electronvolt}$, $a=b_x = b_z=0$ and $E_{v,L}=\SI{76}{\micro\electronvolt}$, $E_{v,R}=\SI{58}{\micro\electronvolt}$, (a) $\vartheta=0$, (b) $\vartheta=\pi/6$, (c) $\vartheta=\pi/2$. In the limit $\vartheta\to 0$ ($\vartheta \to \pi/2$) the valley-flip (valley-conserving) tunneling matrix elements $\propto \sin\vartheta$ ($\propto\cos\vartheta$) vanish. Pink and green ellipses in (b) indicate where spin-conserving valley transitions allow for two interfering paths for both spin states. Arrows indicate upper and lower paths for the spin ground state in the excited valley which splits at $A1$ and can interfere at $A2$.\label{fig_ValleySpec}}
\end{figure}
The new basis states are $\{|\sigma v,0\rangle,|0,\sigma v\rangle\}$ with $v\in\{1,2\}$ indicating the local valley eigenstate and $\sigma \in\{\uparrow,\downarrow\}$ the spin state. The valley index $v=1$ is chosen to denote the valley ground state. An example level diagram of $H_v$ for the cases $\vartheta = 0,\pi/6,\pi/2$ is plotted in Fig.~\ref{fig_ValleySpec}.

In addition to the charge and spin fidelities $F_c$ and $F_s$, we now introduce the spin-valley fidelity $F_{s,v}=|\langle 0,\sigma v|\mathrm{out}\rangle|^2$ to quantify to which degree the protocol transports information encoded in spin and valley simultaneously. For $0<\vartheta <\pi$, the local valley bases are not collinear and valley-flip tunneling occurs. In the case $\vartheta = \pi/2$ every tunneling event flips the valley quantum number.

\subsection{Valley-induced charge errors\label{sec_ValleyChargeError}}

\begin{figure}
\includegraphics{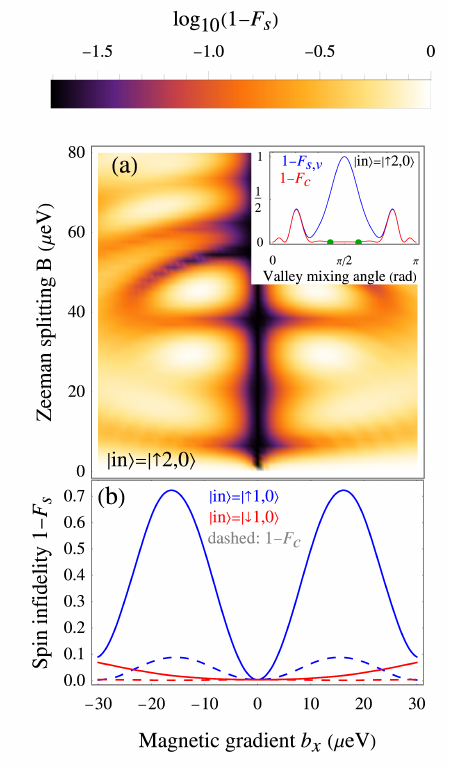}%
\caption{Spin infidelity with valley degree of freedom.
(a) Plotted is the logarithm of the spin infidelity $1-F_s$ as a function of the magnetic field $B$ and the transverse magnetic field gradient $b_x$ for $|\mathrm{in}\rangle = |\!\uparrow 2,0\rangle$ with $a=b_z=0$, $t_c = \SI{21}{\micro\electronvolt}$, $\eps_0=\SI{0.8}{\milli\electronvolt}$ and $E_{v,L}=\SI{76}{\micro\electronvolt}$, $E_{v,R}=\SI{58}{\micro\electronvolt}$, $\vartheta=\pi/4$. Interference extrema of first and second order due to both, spin and valley transitions are present. The fine ripples are finite-time LZ effects due to the relatively short ramp.
(Inset) Spin-valley infidelity $1-F_{s,v}$ (blue) and charge infidelity $1-F_c$ (red) as function of the valley mixing angle $\vartheta$ for $|\mathrm{in}\rangle = |\!\uparrow 2,0\rangle$ and parameters as in the main plot, $B=\SI{54}{\milli\electronvolt}$, $b_x=0$. At $\vartheta=\pi/2$ the valley-conserving tunneling matrix elements vanish. The transport of spin and valley is limited by the charge shuttling infidelity $1-F_c$ whose maxima are due to destructive LZ interference suppressing the charge transfer. Green dots indicate spots with vanishing infidelity for which closed analytic expressions were found in Sec.~\ref{sec_AnalyticalModel}.
(b) Spin (charge) infidelity plotted as solid (dashed) lines for $|\mathrm{in}\rangle = |\uparrow 1,0\rangle$ (blue) and $|\mathrm{in}\rangle = |\downarrow 1,0\rangle$ (red) as a function of the transverse magnetic gradient $b_x$ with $B=\SI{42}{\micro\electronvolt}$, remaining parameters as in (a). In the case $|\mathrm{in}\rangle = |\!\uparrow 1,0\rangle$ a large spin shuttling error may come from a transition to $|\mathrm{out}\rangle = |0,\downarrow 2\rangle$ while the charge is transported faithfully.\label{fig_ValleyFidelity}}
\end{figure}

The additional avoided crossings allow for a large number of paths that can lead to faithful spin transport. Even in the absence of SOI and magnetic gradients, spin-preserving transitions between different valley states can lead to LZ interference between the paths if the electron is initialized in the excited valley state, similar to the interference observed for the excited spin state. Destructive interference can lead to a spin-conserving transition into the excited charge state with the opposite valley quantum number instead of a charge transfer. Since no spin transition is involved, both spin states are affected equally by a change of valley parameters. Figure \ref{fig_ValleyFidelity}a shows the LZ interference extrema of $1-F_s$ of first and second order due valley transitions and spin transitions in the $b_x$-$B$-plane for a spin prepared in the excited valley state $|\mathrm{in}\rangle = |\!\uparrow 2,0 \rangle$.

The angle $\vartheta$ parametrizes the ratio of valley-flipping to valley-conserving tunneling terms, similar to $b_x$ in the case of magnetic fields. In direct analogy, $\vartheta = 0$ and $\vartheta = \pi$ correspond to $b_x=0$ while $\vartheta =\pi/2$ corresponds to $b_x\to \infty$. Consequently, the charge infidelity $1-F_c$ (which lower-bounds $1-F_s$ and $1-F_{s,v}$) can be drastically enhanced for certain values of $\vartheta$, as the inset of Fig.~\ref{fig_ValleyFidelity} shows. This is in analogy with the spin-related interference extrema in Fig.~\ref{fig_SpinFids}b. The minimal error at $\vartheta \in\{0,\pi\}$ is easily explained by the fact that there is no valley-flip tunneling in this case and the corresponding crossings in the spectrum remain closed. Analogously, at $\vartheta = \pi/2$ there are no valley-conserving transitions and again no interfering paths can be found, as emphasized by Fig.~\ref{fig_ValleySpec}a,b.

In an experimental realization it is challenging to control the valley pseudospin. The valley can be faithfully initialized to $|\mathrm{in}\rangle =| \sigma 1,0\rangle$ by relaxation. The results of a shuttling protocol for $\sigma = \uparrow, \downarrow$ are displayed in Fig.~\ref{fig_ValleyFidelity}b. If $|\mathrm{in}\rangle = |\!\uparrow 1,0\rangle$ interference due to spin transitions as discussed in Sec.~\ref{sec_SingleShuttles} can still occur. However, with non-vanishing valley-flip tunneling there is an additional path leading to the transition $|\!\uparrow 1,0\rangle \to |0,\downarrow 2\rangle$. In this case the spin information is lost while the charge is still transported with relatively low infidelity. If both, spin and valley are initialized to the ground state no interference can occur. The shuttling fidelity is limited by LZ transitions due to non-collinear local bases in analogy to the case $|\mathrm{in}=|\!\downarrow,0\rangle$ without regard of the valley.

As a function of the valley splittings $E_{v,L}$ and $E_{v,R}$ the infidelity shows an oscillating pattern which agrees with previous results \cite{HuShuttling}. This can be seen in Fig.~\ref{fig_ValleySplittingPlane} for one particular choice of $\vartheta$ and $B$ comparable to $E_{v,L(R)}$.
If one valley splitting is sufficiently smaller than the orbital splitting $2t_\mathrm{sc}$ only one avoided crossing between different valley states can form and no interference is observed. If both $E_{vi}\ll 2t_\mathrm{sc},\ i\in\{L,R\}$ there are no anticrossings between valley states at all. An analogous comparison with the Zeeman splitting $B$ can be made \cite{HuShuttling}.

\subsection{Analytical model\label{sec_AnalyticalModel}}

To gain further insight into the mechanism behind the charge error reported in Sec.~\ref{sec_ValleyChargeError} an analytical model is derived to estimate the infidelity. To that end, a first-order Schrieffer-Wolff (SW) transformation \cite{SW,SW_math} is applied to find effective two-level Hamiltonians for the anticrossings $A_1$ at $\eps= \frac{1}{2}(B_R - B_L) -E_{v,L}$ and $A_2$ at $\eps= \frac{1}{2}(B_R - B_L) + E_{v,R}$ indicated in Fig.~\ref{fig_ValleySpec}b. These are the first and the last avoided crossing between opposite valley states passed in the shuttling protocol for $|\mathrm{in}\rangle = |\!\downarrow 2,0\rangle$. The same can be done at the respective avoided crossings for the excited spin state. This reduces the Hamiltonian $H_v$ to a LZ problem
\begin{equation}
H_{Aj} = x_j \mathbbm{1} +
\left(
\begin{array}{cc}
(\eps+z_j)/2 & y_j^* \\
 y_j & -(\eps+z_j)/2 \\
\end{array}
\right)\label{eq_HEff-Valley}
\end{equation}
for each anticrossing $j\in\{1,2\}$ individually. Thus, the LZ formula can be applied to obtain transition probabilities
\begin{equation}
P_j = e^{-2\pi |y_j|^2/\alpha},\label{eq_PLZ}
\end{equation}
at $A_j$. Additionally, the Stokes phase $\varphi_j$ associated with each avoided crossing is computed \cite{AnnPhys210.16,PhysRep492.1},
\begin{equation}
\varphi_j = \frac{|y_j|}{\alpha}\left(\ln\frac{|y_j|}{\alpha}-1\right) + \mathrm{arg}\:\Gamma\left(1-i \frac{|y_j|}{\alpha}\right)-\frac{\pi}{4},
\end{equation}
where $\Gamma$ is the Gamma function. 

For the avoided crossings between $A1$ and $A2$ it is assumed that the passage through any anticrossing of states with opposite spin is perfectly diabatic while anticrossings of states with opposite charge but same spin state are assumed to be passed adiabatically. These assumptions are justified by the observation of highly adiabatic charge and spin transfer (Fig.~\ref{fig_SpinFids}) far from the LZ interference due to spin transitions. Consequently, the approximation is only valid for negligible spin-flip tunneling $t_\mathrm{sf}\ll t_\mathrm{sc}$. This procedure identifies two possible paths for each spin state. The phase difference
\begin{equation}
\Delta \varphi = \varphi_s -\frac{1}{\alpha} \int_{A_1}^{A_2} \mathrm{d}\eps \left(E_-(\eps) - E_+(\eps)\right)
\end{equation}
between the two paths includes the dynamical phase difference and the Stokes phases $\varphi_s$ associated with the avoided crossings along the paths.

Starting with the state $|\mathrm{in}\rangle = |\sigma 2,0\rangle$ the probability of a charge error to occur is given as
\begin{eqnarray}
P_\mathrm{err} &=& P_1 + P_2 - 2 P_1 P_2\label{eq_pErr}\\
&&+ 2 \sqrt{(1 - P_1) P_1 (1 - P_2) P_2} \cos\left(\varphi_1 + \varphi_2 + \Delta\varphi\right),\nonumber
\end{eqnarray}
where the $\cos$-term describes LZ interference. To calculate the dynamical phase difference $\Delta\varphi$ in the vicinity of an avoided crossing at position $\eps = -z$ with interaction term $y$ the adiabatic states are approximated as functions $E_\pm(\eps)\approx \pm\sqrt{(\eps + z)^2+4|y|^2}$. The minima of $P_\mathrm{err}=1-F_c$ coincide with a good accuracy with the numerical simulation of the Schr\"odinger equation. Alternatively, the eigenstates can be integrated numerically to avoid approximations when computing $\Delta\varphi$. This approach yields slightly better results but cannot give analytical solutions.

\begin{figure}
\includegraphics{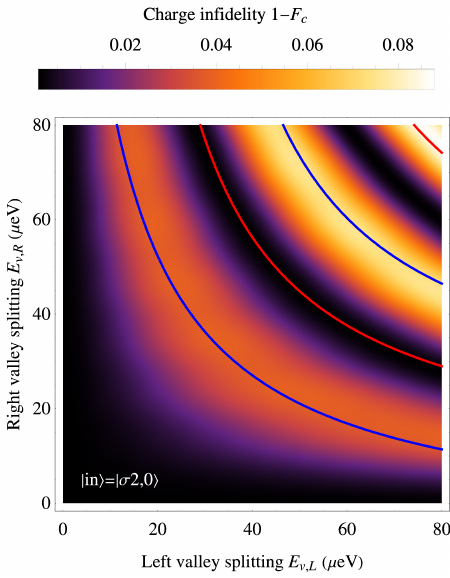}%
\caption{Charge infidelity $1-F_c$ for $|\mathrm{in}\rangle =|\sigma 2,0\rangle$ as a function of $(E_{v,L},E_{v,R})$ for $B=\SI{54}{\micro\electronvolt}$, $a=b_z=b_x=0$, $t_c = \SI{21}{\micro\electronvolt}$ $\vartheta=\pi/4$ and $\eps_0=\SI{8}{\milli\electronvolt}$. The density plot is a numerical result, blue (red) solid lines indicate maxima (minima) predicted by the analytical model after a constant shift $E_{v,j}\mapsto E_{v,j}+k$. For small valley splitting the agreement is good, however, the second order minimum of $1-F_s$ has a large deviation already, as seen in the upper right corner.\label{fig_ValleySplittingPlane}}
\end{figure}

After the SW transformation the valley splittings enter Eq.~(\ref{eq_pErr}) only via the phases $\varphi_j$ and $\Delta \varphi$. Thus, the dependence on the valley splittings is only due to the interference term. The extrema of the error probability in the $E_{v,L}$-$E_{v,R}$-plane are determined by the relation $\sin\left(\varphi_1 + \varphi_2 + \Delta\varphi\right)=0$. No solution in closed form could be found, therefore we numerically solve this equation and find that the solution can be fitted to the contour $(E_{v,L}-k_1)(E_{v,R}-k_1)\approx k_2$ with high accuracy. This agrees with the numerics up to a variation in the shift $k_1$. As shown in Fig.~\ref{fig_ValleySplittingPlane}, the analytically derived maxima and minima of this interference pattern agree well with the first few orders of numerically computed extrema up to a constant shift in the $E_{v,L}$-$E_{v,R}$-plane, $E_{v,j}\mapsto E_{v,j}+k$.  For large $E_{v,j}$ the approximations in the analytical model lead to deviations. This is particularly relevant for higher order extrema.

Besides the adiabatic protocol $\lim_{\alpha\to 0}P_\mathrm{err}=0$ a number of minima of $P_\mathrm{err}$ can be found as a closed expression, given in Appendix~\ref{app_DetailsOnModel} by Eq.~(\ref{Eq_magicTh1})-(\ref{Eq_magicTh2}). 
In the inset of Fig.~\ref{fig_ValleyFidelity} these solutions are indicated by green dots. As a function of  $\vartheta$ these minima frame an interval centered around the point $\vartheta = \pi/2$ with minimal infidelity. It is worth noting that these solutions are independent of the valley splittings $E_{v,j}$. If the Zeeman splitting becomes dominant over the tunnel splitting the assumptions of the model are no longer valid and the equations yield no solution.

At $\vartheta = 0,\pi$ the valley-flipping matrix elements vanish, $y_1 = y_2 =0$, and the avoided crossings remain closed. Consequently, the LZ probabilities are $P_1=P_2 = 1$ resulting in $P_\mathrm{err}=0$. Both $y_j$ reach their maximum at $\vartheta=\pi/2$. Due to the dependence given by Eq.~(\ref{eq_PLZ}) the probability exponentially decays to its minimum value at $\vartheta=\pi/2$, granting a plateau with low infidelity due to the tail of the exponential function. In the regime $0 < \vartheta < \pi/2$ between these two limiting cases the infidelity can also fall to minimal values, ensured by constructive interference depending on $E_{v,i}$. 
%Details can be found in App.~\ref{app_DetailsOnModel}.

A physical process that can lead to dephasing and thus reduce the effectiveness of constructive interference is charge noise \cite{doi:10.1063/1.4954700,PhysRevB.91.235411,PhysRevB.94.165411,Russ_2017,Hooge1997,Taylor2006}. 
%In Fig.~\ref{fig_ValSplitDep} the dephased error probability %can be compared to $P_\mathrm{err}$, Eq.~(\ref{eq_pErr}). 
We find that in the presence of charge noise both the maxima and minima of $1-F_c$ collapse to the dephased function $P_\mathrm{err}^\mathrm{deph} = P_1 + P_2 - 2 P_1 P_2$. 
Furthermore, in the literature there are plenty of results already dealing with the LZ problem under the influence of charge noise \cite{PhysRevA.91.052103,PhysRevB.87.165425,PhysRevB.87.224301%,PhysRevB.76.024310,PhysRevLett.110.086804,PhysRevB.67.144303,ChargeNoise
} which can be adopted to estimate the changes in the transition probability at the avoided crossings. 

Note that even though this model was derived to analyze the interference between valley states while the spin is transported faithfully it can easily be adapted to the interference between spin states in the local eigenbasis in the case of irrelevant valley splittings %$E_{v,j}\gg B$ by simply swapping valley and spin.
This is accomplished by mapping $E_{v,j}\to B_j$ and $\tan\vartheta$ to the ratio $t_\mathrm{sf}/t_\mathrm{sc}$.

Respecting both, spin and valley, it is in principle possible to achieve a spin infidelity which compares well to the case without valley. However, since the valley parameters are set during device fabrication and are often not controllable to a large degree an unfortunate occurrence can significantly increase the infidelity compared to the case with spin only. It is still possible to exploit constructive interference to achieve a low shuttling infidelity, although this requires precise tuning of the device parameters, which may be difficult to achieve in practice.

\section{Conclusions\label{sec_conclusion}}

A detuning sweep in a singly occupied DQD was investigated as a building block of a scalable electron shuttling protocol. It was found that although an infidelity of $10^{-5}$ of charge transfer, even without optimization of the pulse shape is realistic, the SOI and magnetic gradients can introduce errors into the spin shuttling. %An engineered transverse magnetic field difference between the dots can compensate errors due to SOI, and vice versa.
Optimal shuttling results are achieved when the spin-flip tunneling term $t_\mathrm{sf}$ vanishes. Transient effects can be mitigated by preparing and measuring stationary local eigenstates of the system and by using a sufficiently large final and initial detuning% to isolate the dots
.

With optimized parameters we find that a spin shuttling infidelity $1-F_s\lesssim 0.002$ for the excited spin state and $1-F_s\lesssim 0.001$ for the spin ground state is achievable for a ramp with a level velocity of $\alpha = \SI{600}{\micro\electronvolt\per\nano\second}$. Realistically, lower fidelities can be expected since a flawless parameter setting would also require control over the valley pseudospin. 
We expect that the numbers reported here can be further improved by optimizing the pulse shape away from a simple ramp \cite{PhysRevLett.116.230503,PhysRevX.7.011021,Buonacorsi2020}.% and minimizing undesired effects by smoothly matching the time dependent and constant gate voltage.

In a periodically driven DQD performing a shuttling sequence the resulting infidelity is modulated by LZ interference due to the repeated passage through the avoided crossings. Diabatic transitions to the other eigenstates during the repeated shuttling protocol are a major loss mechanism %establishing a steady state with equal population in all available states 
in a fast protocol.
We examined the interaction with nearby reservoirs in long shuttling sequences as an example how the environment can lead to a significant shuttling error probability. Incoherent spin-flip cotunneling between the system and the reservoirs further increases the spin infidelity and also introduces decoherence limiting the observation of interference effects.% For a small number of shuttles the reservoirs are negligible.

Due to multiple avoided crossings involving the excited spin or valley state strong transport errors up to complete destructive interference can occur even in single shuttles if the protocol is not sufficiently adiabatic. For $B\gtrsim 2t_c$ and $t_\mathrm{sf}\neq 0$ avoided crossings between opposite spin states open interfering paths. Similarly, when $E_{v,L},E_{v,R}\gtrsim 2t_\mathrm{sc},B$ valley transitions become significant.
However, this entails possible applications of LZ interference in a DQD as a filtering or readout device to spatially separate electrons with different spin or valley state. Sophisticated device engineering and control provided, constructive interference can ensure high fidelity shuttling.

The interference allowed by spin-conserving valley transitions was characterized in detail numerically and by means of an approximate analytical model which can also be adopted to the situation of valley-conserving spin transitions. The analysis confirms that destructive LZ interference is a major error mechanism in fast electron shuttling protocols and provides a means to estimate optimal experimental regimes where the transport fidelity is protected by constructive interference. The analytical model is still not perfectly precise. Possibly, further improvements could be achieved by using the full finite-time solution of the LZ problem \cite{PhysRevA.53.4288} instead of the LZ formula.

%Another open task is the description and optimization of spin shuttling in more complex systems to understand possibilities and limitations due to geometry and coupling to the environment.

%%%%%%%%%%%%%%%%%%%%%%%%%%%%%%%%%%%%%%%%%%%%%%%%%%%%%%%%%%%%%%%%%
%%%%%%%%%%%%%%%%%%%%%%%%% Appendices %%%%%%%%%%%%%%%%%%%%%%%%%%%%%%
%%%%%%%%%%%%%%%%%%%%%%%%%%%%%%%%%%%%%%%%%%%%%%%%%%%%%%%%%%%%%%%%%

\appendix

\section{Magnetic field contributions\label{app_MagField}}

Here, we detail the different contributions to the Zeeman terms in $H_0$ and $H_\mathrm{grad}$, Eqs.~(\ref{eq_H0}) and (\ref{eq_Hgrad}). Our Hamiltonian includes a homogeneous external magnetic field $\vec{\tilde{B}}_\mathrm{ext}$, the field of a micromagnet, $\vec{\tilde{B}}_m^{L(R)}$, which is inhomogeneous and thus different in the left and right dot and lastly the hyperfine interaction. The tilde  denotes the use of magnetic field rather than energy units.  Although the hyperfine interaction can be suppressed by choosing isotopically purified silicon as host material \cite{EnrichedSilicon} in natural silicon only $\approx 95\%$ of the nuclei are non-magnetic \cite{RevModPhys.79.1217}. For the purpose of this work a semiclassical description of the hyperfine interaction with the Overhauser field $\vec{\tilde{B}}_N^i=(g_i \mu_B)^{-1}\sum_k A_k \vec{I}_k$ is sufficient \cite{RevModPhys.79.1217}. Here, $g_i$ is the electron $g$-factor in dot $i=L,R$, $A_k$ is the coupling between the electron and the nuclear spin $\vec{I}_k$ and $\mu_B$ is the Bohr magneton.

With $\vec{\tilde{B}}_\mathrm{int}^i= \vec{\tilde{B}}_m^i + \vec{\tilde{B}}_N^i$ the Zeeman Hamiltonian in dot $i$ is given by $H_z^i=g_i \mu_B (\vec{\tilde{B}}_\mathrm{ext}+ \vec{\tilde{B}}_\mathrm{int}^i)\cdot \vec{S}$ where $\vec{S}=\hbar \boldsymbol{\sigma}/2$ is the electron spin operator. With $\vec{B} = \hbar\mu_B\left[g_L \left(\vec{\tilde{B}}_\mathrm{ext}+\vec{\tilde{B}}_\mathrm{int}^L\right)+g_R \left(\vec{\tilde{B}}_\mathrm{ext}+\vec{\tilde{B}}_\mathrm{int}^R\right)\right]/2$ and $\vec{b} = \hbar\mu_B\left[g_L \left(\vec{\tilde{B}}_\mathrm{ext}+\vec{\tilde{B}}_\mathrm{int}^L\right)-g_R \left(\vec{\tilde{B}}_\mathrm{ext}+\vec{\tilde{B}}_\mathrm{int}^R\right)\right]/2$ we find the total Zeeman Hamiltonian $H_z^\mathrm{tot} = H_z^L \oplus H_z^R = \left( \mathbbm{1} \otimes \vec{B}\cdot\boldsymbol{\sigma} +  \tau_z \otimes \vec{b}\cdot\boldsymbol{\sigma}\right)/2$. The homogeneous part of $H_z^\mathrm{tot}$ is included into $H_0$ while the inhomogeneous part is treated as $H_\mathrm{grad}$.

\section{Terms of the master equation\label{app_ME}}

Using the notation, $(0,0)=0,\ (\uparrow, 0)=1,\ (\downarrow,0)=2,\ (0,\uparrow)=3,\ (0,\downarrow)=4\, (\downarrow,\downarrow)=5,\ (\uparrow,\downarrow)=6,\ (\downarrow,\uparrow)=7,\ (\uparrow,\uparrow)=8$ for the spin and charge configurations the terms $\gamma_{nm}$ in Eq.~(\ref{eq_ME-res}) are given by
\begin{eqnarray}
\gamma_{nn} &=& \sum_l w_{ln},\\
\gamma_{12} &=& \frac{1}{2}(w_{01}+w_{02}+w_{61}+w_{81}+w_{52}+w_{72}+w_{21}),\\
\gamma_{13} &=& \frac{1}{2}(w_{01}+w_{03}+w_{61}+w_{81}+w_{73}+w_{83}+w_{21}+w_{43}),\nonumber\\
\\
\gamma_{14} &=& \frac{1}{2}(w_{01}+w_{04}+w_{61}+w_{81}+w_{54}+w_{64}+w_{21}),\\
\gamma_{23} &=& \frac{1}{2}(w_{02}+w_{03}+w_{52}+w_{72}+w_{73}+w_{83}+w_{43}),\\
\gamma_{24} &=& \frac{1}{2}(w_{02}+w_{04}+w_{52}+w_{72}+w_{54}+w_{64}),\\
\gamma_{34} &=& \frac{1}{2}(w_{03}+w_{04}+w_{73}+w_{38}+w_{54}+w_{64}+w_{43}).
\end{eqnarray}
Furthermore, the conjugate terms $\gamma_{21}$, $\gamma_{31}$, $\gamma_{41}$, $\gamma_{32}$, $\gamma_{42}$, $\gamma_{43}$ occur with respectively interchanged indices. The contributions $w_{21}=1/T_{1L}$ and $w_{43}=1/T_{1R}$ account for spin relaxation in the left and right dot.

\section{Minima of the error probability\label{app_DetailsOnModel}}

Some minima of the infidelity as a function of $\vartheta$ can be derived analytically. Introducing the notations $b=b_x+ib_y$ and $\tilde\tau= t_c+is$, we used the following conditions to obtain the local minima of $P_\mathrm{err}(\vartheta)=1-F_c(\vartheta)$,
\begin{eqnarray}
\vartheta &=& (-1)^k \arcsin\Big[\Big((  b_z-B) (B + b_z)^2 (a b + (B - b_z) \tilde\tau^*) \nonumber\\
&& + b^* ((B^2 - b_z^2) (b \tilde\tau - (B - b_z) a^*) + b (B^2 + b_z^2) \tilde\tau^*)\Big) \nonumber\\
&&  \Big(-B \tilde\tau^* (2 B |a|^2 + 4 B |\tilde\tau|^2 + 2 i \mathrm{Re}(a b) s)\Big)^{-1}\Big]^{1/2} \nonumber\\
&& + n \pi , \label{Eq_magicTh1}
\end{eqnarray}
\begin{eqnarray}
\vartheta &=&\arctan\Big\{ (-1)^k
\Big[ b^* (b \tilde\tau b_z^2 - (B^2 - b_z^2)((B + b_z)a^* - b t_c) ) \nonumber\\
&& - (B - b_z)^2 (B + b_z) (a b + 2 \tilde\tau (B + b_z)) \Big]^{1/2}
\Big[ (B - b_z)^2 \nonumber\\
&& (B + b_z) (a b + 2 \tilde\tau (B + b_z)) - 8 B^2 \tilde\tau |\tilde\tau|^2 + i b (2 a B \tilde\tau \nonumber\\
&& - b_z^2 b^*)s + a^* (b^* ((B - b_z)(B + b_z)^2 + 2 i B \tilde\tau s) - 4 a B^2 \tilde\tau) \nonumber\\
&& - |b|^2 B^2 t_c \Big]^{-1/2}
\Big\}
+2\pi  n,\label{Eq_magicTh2}
\end{eqnarray}
with $n\in\mathbbm{N}_0$, and $k\in\{0,1\}$. These solutions are %plotted in Fig.~\ref{fig_ZeroErrValAngle} and 
indicated in the inset of Fig.~\ref{fig_ValleyFidelity}. Note however, that these are not all minima of $P_\mathrm{err}(\vartheta)$.

%%%%%%%%%%%%%%%%%%%%%%%%%%%%%%%%%%%%%%%%%%%%%%%%%%%%%%%%%%%%%%%%%
%%%%%%%%%%%%%%%%%%%%%%%% Post-Appendix %%%%%%%%%%%%%%%%%%%%%%%%%%%%%
%%%%%%%%%%%%%%%%%%%%%%%%%%%%%%%%%%%%%%%%%%%%%%%%%%%%%%%%%%%%%%%%%

\begin{acknowledgments}
We thank Maximilian Russ, Amin Hosseinkhani, Hugo Ribeiro, M$\acute{\mbox{o}}$nica Benito, Benjamin D'Anjou, and Philipp Mutter for helpful discussions. Florian Ginzel acknowledges a scholarship from the Stiftung der Deutschen Wirtschaft (sdw) which made this work possible. This work has been supported by ARO grant number W911NF-15-1-0149.
\end{acknowledgments}

\bibliography{Shuttling_lit.bib}

\end{document}